\documentclass[aps,prl,twocolumn,superscriptaddress,amsfonts,amsmath,amssymb,showpacs,floatfix,nofootinbib]{revtex4-1}

\usepackage{graphicx}
\usepackage{longtable}
\usepackage{bm}
\usepackage{multirow}
\usepackage{hyperref}
\usepackage{ulem}
\hypersetup{colorlinks,citecolor=blue,filecolor=blue,linkcolor=blue,urlcolor=blue}
\usepackage{color}
\usepackage{dcolumn}

\begin{document}

\title{First Exploration of Neutron Shell Structure Below Lead and Beyond {\boldmath$N = 126$}}

\author{T.~L.~Tang}
\affiliation{Physics Division, Argonne National Laboratory, Argonne, Illinois 60439, USA}
\author{B.~P.~Kay}
\email{kay@anl.gov}
\affiliation{Physics Division, Argonne National Laboratory, Argonne, Illinois 60439, USA}
\author{C.~R.~Hoffman}
\affiliation{Physics Division, Argonne National Laboratory, Argonne, Illinois 60439, USA}
\author{J.~P.~Schiffer}
\affiliation{Physics Division, Argonne National Laboratory, Argonne, Illinois 60439, USA}
\author{D.~K.~Sharp}
\affiliation{Department of Physics, University of Manchester, M13 9PL Manchester, United Kingdom}
\author{L.~P.~Gaffney}
\affiliation{CERN, CH-1211 Geneva, Switzerland}
\author{S.~J.~Freeman}
\affiliation{Department of Physics, University of Manchester, M13 9PL Manchester, United Kingdom}
\author{M.~R.~Mumpower}
\affiliation{Theoretical Division, Los Alamos National Laboratory, Los Alamos, New Mexico, 87545, USA}
\affiliation{Center for Theoretical Astrophysics, Los Alamos National Laboratory, Los Alamos, NM, 87545, USA}
\author{A.~Arokiaraj}
\affiliation{KU Leuven, Intituut voor Kern-en Stralingsfysica, Celestijnenlaan 200D, 3001 Leuven, Belgium}
\author{E.~F.~Baader}
\affiliation{CERN, CH-1211 Geneva, Switzerland}
\author{P.~A.~Butler}
\affiliation{Oliver Lodge Laboratory, University of Liverpool, L69 7ZE Liverpool, United Kingdom}
\author{W.~N.~Catford}
\affiliation{Department of Physics, University of Surrey, Guildford GU2 7XH, United Kingdom}
\author{G.~de~Angelis}
\affiliation{Istituto Nazionale di Fisica Nucleare, Laboratori Nazionali di Legnaro, I-35020 Legnaro, Italy}
\author{F.~Flavigny}
\affiliation{Institut de Physique Nucl\'{e}aire, CNRS-IN2P3, Universit\'{e} Paris-Sud, Universit\'{e} Paris-Saclay, 91406 Orsay, France}
\affiliation{LPC Caen, Normandie Universit\'{e}, ENSICAEN, UNICAEN, CNRS/IN2P3, 14000 Caen, France}
\author{M.~D.~Gott}
\affiliation{Physics Division, Argonne National Laboratory, Argonne, Illinois 60439, USA}
\author{E.~T.~Gregor}
\affiliation{Istituto Nazionale di Fisica Nucleare, Laboratori Nazionali di Legnaro, I-35020 Legnaro, Italy}
\author{J.~Konki}
\affiliation{CERN, CH-1211 Geneva, Switzerland}
\author{M.~Labiche}
\affiliation{Nuclear Physics Group, UKRI-STFC Daresbury Laboratory, Daresbury, Warrington WA4 4AD, United Kingdom}
\author{I.~H.~Lazurus}
\affiliation{Nuclear Physics Group, UKRI-STFC Daresbury Laboratory, Daresbury, Warrington WA4 4AD, United Kingdom}
\author{P.~T.~MacGregor}
\affiliation{Department of Physics, University of Manchester, M13 9PL Manchester, United Kingdom}
\author{I. Martel}
\affiliation{Oliver Lodge Laboratory, University of Liverpool, L69 7ZE Liverpool, United Kingdom}
\author{R.~D.~Page}
\affiliation{Oliver Lodge Laboratory, University of Liverpool, L69 7ZE Liverpool, United Kingdom}
\author{Zs.~Podoly\'{a}k}
\affiliation{Department of Physics, University of Surrey, Guildford GU2 7XH, United Kingdom}
\author{O.~Poleshchuk}
\affiliation{KU Leuven, Intituut voor Kern-en Stralingsfysica, Celestijnenlaan 200D, 3001 Leuven, Belgium}
\author{R.~Raabe}
\affiliation{KU Leuven, Intituut voor Kern-en Stralingsfysica, Celestijnenlaan 200D, 3001 Leuven, Belgium}
\author{F. Recchia}
\affiliation{Dipartimento di Fisica e Astronomia, Università degli Studi di Padova, I-35131 Padova, Italy}
\affiliation{INFN, Sezione di Padova, I-35131 Padova, Italy}
\author{J.~F.~Smith}
\affiliation{SUPA, School of Computing, Engineering, and Physical Sciences, University of the West of Scotland, Paisley PA1 2BE, United Kingdom}
\author{S.~V.~Szwec}
\affiliation{Department of Physics, University of Jyvaskyla, P.O. Box 35, FI-40014 Jyvaskyla, Finland}
\author{J.~Yang}
\affiliation{KU Leuven, Intituut voor Kern-en Stralingsfysica, Celestijnenlaan 200D, 3001 Leuven, Belgium}


\begin{abstract}

The nuclei below lead but with more than 126 neutrons are crucial to an understanding of the astrophysical $r$-process in producing nuclei heavier than $A\sim190$. Despite their importance, the structure and properties of these nuclei remain experimentally untested as they are difficult to produce in nuclear reactions with stable beams. In a first exploration of the shell structure of this region, neutron excitations in $^{207}$Hg have been probed using the neutron-adding ($d$,$p$) reaction in inverse kinematics. The radioactive beam of $^{206}$Hg was delivered to the new ISOLDE Solenoidal Spectrometer at an energy above the Coulomb barrier. The spectroscopy of $^{207}$Hg marks a first step in improving our understanding of the relevant structural properties of nuclei involved in a key part of the path of the $r$-process.

\end{abstract}

\maketitle

\begin{figure*}
\centering
\includegraphics[scale=0.54]{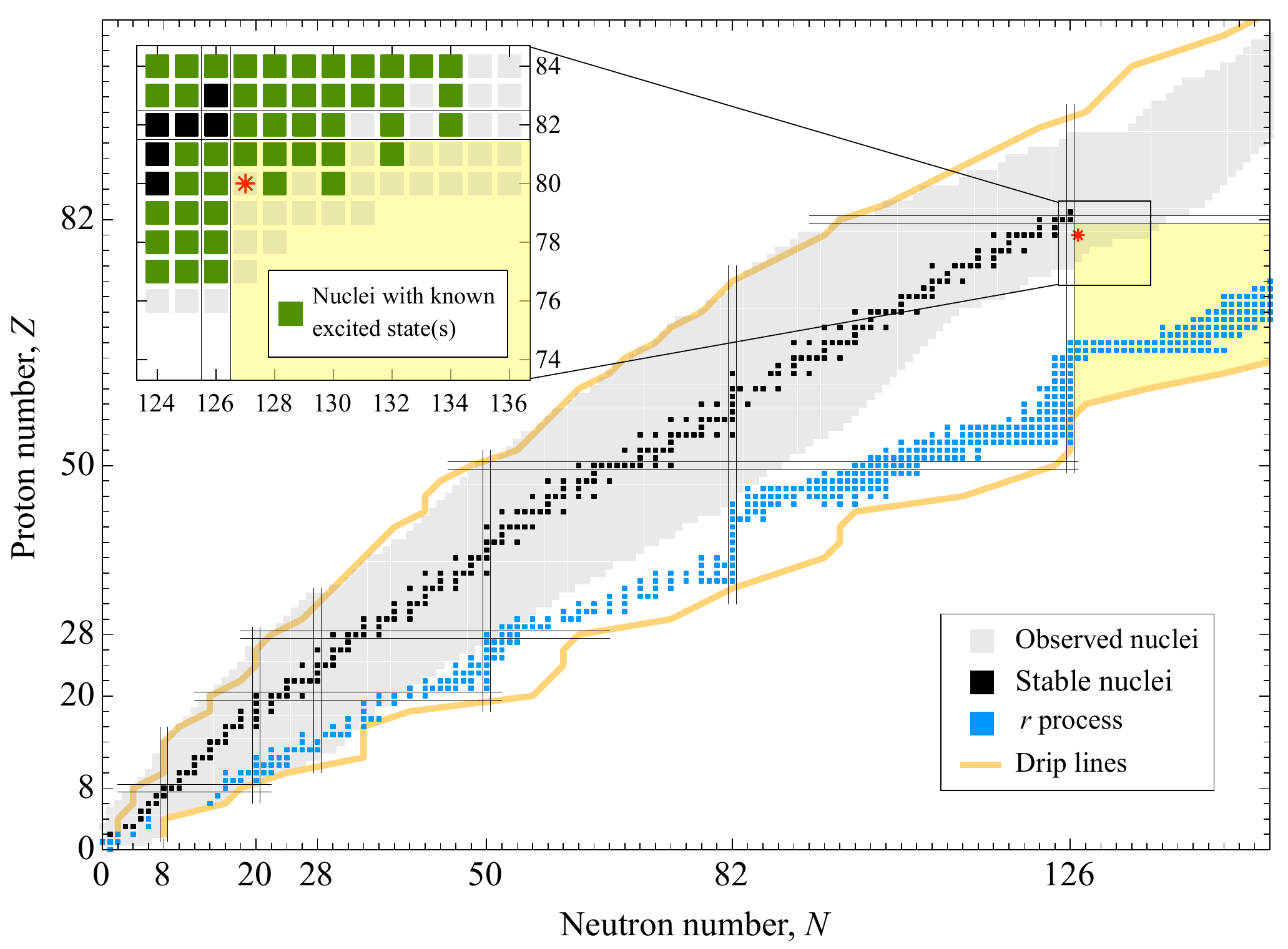}
\caption{\label{fig1} The chart of nuclides color coded to show the stable nuclei (288) in black and nuclei that have been shown to exist in grey ($\sim$3350). Approximately 7000 nuclei, inside the drip lines, are predicted to exist~\cite{Erler12}. Nuclei estimated to be involved in the astrophysical $r$ process~\cite{jina} are shown in blue. The isotope $^{207}$Hg is marked with a red asterisk, lying in the region shaded yellow with boundaries at $Z=82$ and $N=126$. The bold lines show the traditional magic numbers. The inset shows the region around $^{207}$Hg. In the inset only, nuclei with at least one known excited state ($\sim$2100 across the entire chart) are shown in green~\cite{nudat,Steer11}.}
\end{figure*}

The nucleus $^{207}$Hg lies in the almost completely unexplored region of the nuclear chart below proton number 82 and just above neutron number 126, both ``magic'' numbers representing closed shells in the nuclear shell model~\cite{Mayer49}. The doubly-magic nucleus $^{208}$Pb is the cornerstone of this region, a benchmark nucleus in our understanding of the single-particle foundation of nuclear structure. This region, highlighted on the nuclear chart in Fig.~\ref{fig1}, is unique in that its single-particle structure remains unexplored.

The nucleosynthesis of heavy elements via the rapid neutron-capture \mbox{($r$-) process} path~\cite{Burbridge57} crosses this region, as shown in Fig.~\ref{fig1}. The robustness of the $N=126$ neutron shell closure plays a crucial role in the nucleosynthesis of the actinides~\cite{Goriely11,Zhu18,Holmbeck18,Vassh19,Eichler19}. The recent observation of a neutron star merger has provided a new focus of interest~\cite{Abbott17,Watson19}, suggesting a possible astrophysical environment for $r$-process nucleosynthesis~\cite{Thielemann17,Cote18,Horowitz19,Cowan19}. 

Approaching the $r$-process path along the \mbox{$N=126$} isotonic chain from Pb, the binding energies (the degree to which neutrons are bound by the mean-field potential created by the decreasing number of all other nucleons) decrease, eventually crossing zero binding and becoming unbound. Near closed shells, the level density is low, so the usual statistical assumptions of many resonances participating in neutron capture is not valid, and specific nuclear-structure properties become important. Knowledge of ground-state binding energies of nuclei with $N=126+n$ is important in defining the waiting point caused by the $N=126$ closure, the bottleneck which is responsible for the third peak in solar system elemental abundances at nuclear mass \mbox{$A\sim195$~\cite{Anders89}}. The binding energies are critical to how the $r$-process evolves. The energies of ground and excited states have significant consequences for the rate at which direct $s$-, $p$-, (and possibly $d$-) wave neutron-capture ($n$,$\gamma$) reactions proceed~\cite{Mathews83,Goriely97,Mumpower16}. This was discussed recently in the context of the $N=82$ shell closure in Ref.~\cite{Manning19}.

As zero binding is approached, the energies of $s$ orbitals increase less rapidly than those of states with higher angular momenta~\cite{Bohr69}. This behavior has been studied for light nuclei~\cite{Hoffman14,Kay17} and, in the vicinity of $N=126$, it is likely to play an important role in neutron-capture reactions. Direct measurements of properties of nuclei in the $r$-process path in this region will not be possible for many years, if not decades. Experimentally, this region of the nuclear chart has remained largely inaccessible. Fragmentation and isotope separation online, which are the methods of choice at current and next generation radioactive ion beam facilities, only produce low-intensity beams of neutron-rich ($N\geq126$) nuclei below $^{208}$Pb. Techniques that could produce these nuclei with significant yields, such as multi-nucleon transfer, look promising although technological developments are necessary to manipulate the reaction products for spectroscopy~\cite{Hirayama17,Savard19}. New data on $^{207}$Hg marks a first step in the study of these systems.

Only one transfer-reaction study has probed the \mbox{$Z<82$}, $N>126$ region using a long-lived radioactive target of $^{210}$Pb~\cite{Ellegaard76}, providing some limited information on proton-hole states in $^{209}$Tl. Beyond the simple existence of some nuclei, derived from decay studies, only limited knowledge of a few excitations in $^{208-210}$Tl (neighboring Pb isotopes) and $^{208,210}$Hg are known~\cite{nudat,Fornal11}. Most recently, $\gamma$-ray transitions have been seen in $^{211,213}$Tl~\cite{Gottardo19}. In no instance is any direct knowledge of single-neutron structure known. All that has been previously established about $^{207}$Hg, the next even-proton member of the $N=127$ isotonic chain below Pb, is an estimate of its lifetime ($T_{1/2}=2.9$~m), its mass~\cite{Kondev11,Wang16}, and that its ground state decays via $\beta$ decay.

In this work, we report the first study of single-neutron excitations beyond $N=126$ for elements below Pb, achieved using a transfer reaction in inverse kinematics with a radioactive  beam of $^{206}$Hg, accelerated to energies above the Coulomb barrier for collisions with deuterium. This study was made possible by new advances in technology at the ISOLDE radioactive-beam facility at CERN~\cite{Ames05,Goodacre16,Kadi17} and with the development of the ISOLDE Solenoidal Spectrometer (ISS)~\cite{ISSwebsite} based on a technique pioneered at Argonne National Laboratory (ANL). As was demonstrated with HELIOS at ANL~\cite{Lighthall10}, factors of 2--3 improvements in $Q$-value resolution can be achieved in the study of these types of reactions using the solenoidal-spectrometer technique~\cite{Schiffer99,Wuosmaa07}.

In this study, about 6000 protons were identified from the $^{206}$Hg($d$,$p$)$^{207}$Hg reaction. These result from the bombardment of approximately $1.5\times10^{11}$ $^{206}$Hg ions with deuterated polyethylene targets of nominal thickness 165~$\mu$g/cm$^2$ at a rate of around 3 to 8$\times$10$^5$ $^{206}$Hg ions per second and an energy of 1.52~GeV (7.38~MeV/u), the highest energy available at the time.

The preparation of the radioactive Hg ions is described elsewhere~\cite{Marsh18,Wenander10}. The Hg ions (46$^+$) were injected into the REX/HIE-ISOLDE linear accelerator in bursts of width $\sim$800-$\mu$s, every 450~ms. The beam was contaminated with $^{130}$Xe at a level of $<2\%$. The level of the Xe impurity was determined from background runs with the ion-source lasers off. A measurement using a pure $^{130}$Xe beam, at the same MeV/u as the Hg, was carried out to corroborate this. 

\begin{figure}
\centering
\includegraphics[scale=0.9]{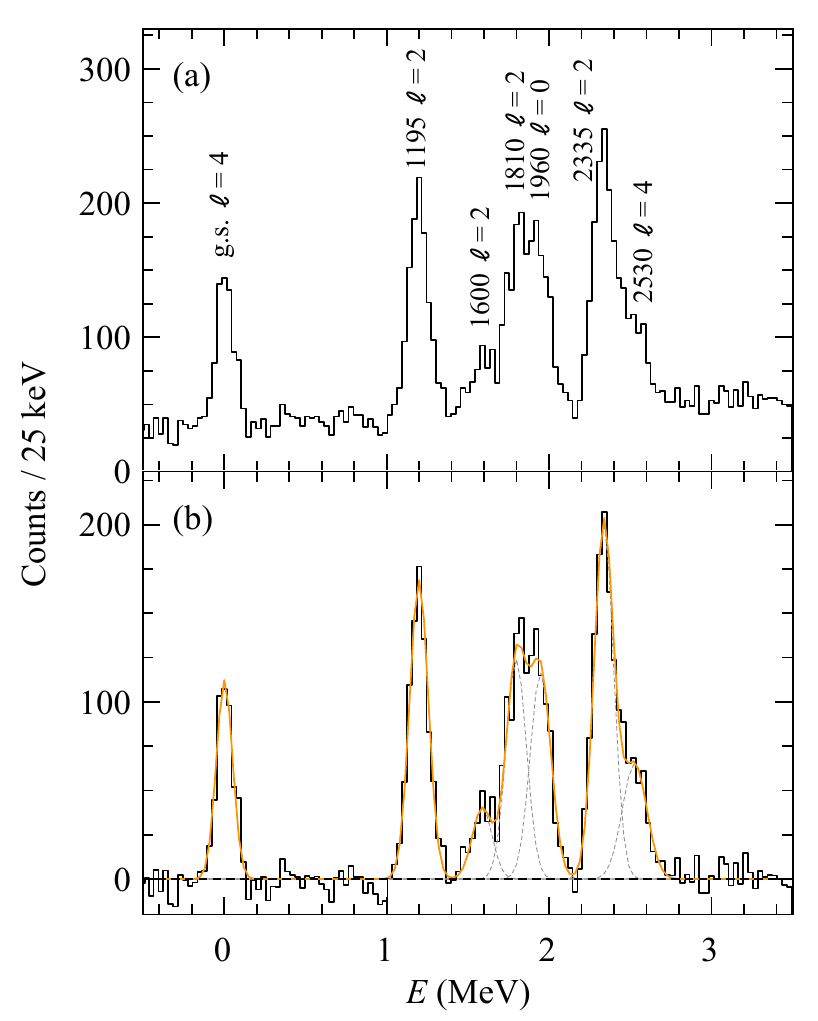}
\caption{\label{fig2} (a) The low-lying excitation-energy spectrum of $^{207}$Hg as measured via the ($d$,$p$) reaction on $^{206}$Hg at 7.38~MeV/u. States are labeled by their energies in keV and $\ell$ value. Plot (b) is the same spectrum with a linear background subtracted and with fits shown~\cite{fits}.}
\end{figure}

\begin{figure}
\centering
\includegraphics[scale=0.9]{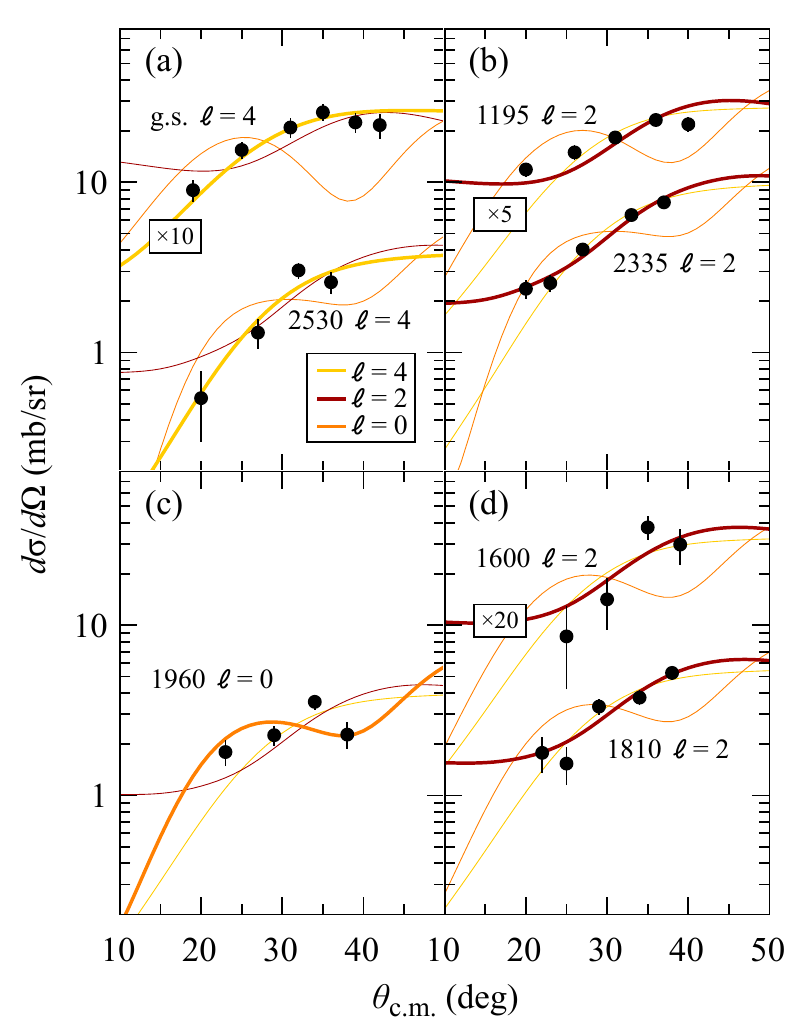}
\caption{\label{fig3} Angular distributions for the outgoing protons from the ($d$,$p$) reaction on $^{206}$Hg at 7.38~MeV/u. The color of the lines correspond to $\ell=0$ (orange), 2 (maroon), and 4 (yellow) transfer. Transitions are labeled by their excitation energies (in keV) in the residual nucleus and the $\ell$-value corresponding to the best fit angular distribution. The thicker curves represent the best fit angular distribution for the $\ell$-value quoted. Thinner lines are used to show other possible $\ell$ values.}
\end{figure}

Thin deuterated polyethylene targets are susceptible to damage even with relatively low ($\ll 10^9$~pps) intensity beams. The damage is reasonably well characterized for light and medium mass beams~\cite{Rehm98,Kay11} but not for heavier beams. In the experiment described here, the target was replaced every 4 hours, on average, resulting in the use of 20 targets in the 82-hour duration. No evidence of target degradation was observed via elastically scattered deuterons and carbon ions in a monitor detector. 

The ISS, in this first stage of its development, was set up in a manner similar to that described in Ref.~\cite{Kay11} using the HELIOS detector system. A 2.5-T magnetic field was used. The data acquisition was triggered by signals in the on-axis Si array~\cite{Lighthall10}. A signal from the REXEBIS was used to gate the data acquisition system during the beam release, disabling it during the charge-breeding process. This suppressed spurious background, such as $\alpha$ decays present due to nuclei produced in fusion-evaporation reactions.

The excitation-energy spectrum for $^{207}$Hg is shown in Fig.~\ref{fig2} both with and without a background subtraction. The background is predominately prompt protons from fusion-evaporation reactions of the beam and $^{12}$C in the target. The $Q$-value resolution was $\sim$140~keV FWHM.  The spectrum represents a sum of all detectors on the Si array which corresponds to center-of-mass angles, $20^{\circ}\lesssim\theta_{\rm cm}\lesssim40^{\circ}$. 

From the data, seven states have been observed below 3~MeV, which are associated with adding a neutron in the vacant $1g_{9/2}$, $2d_{5/2}$, $3s_{1/2}$, $2d_{3/2}$, and $1g_{7/2}$ orbitals beyond $N=126$, as shown in Table~\ref{tab1}. The high-$j$ $0i_{11/2}$ $\ell=6$ strength, estimated to lie at around 0.8~MeV, is not seen; at an incident beam energy of 7.38~MeV/u, ($d$,$p$) yields corresponding to $\ell=6$ transfer are expected to be $<$10 counts (for a pure single-particle state) in total. The negative parity $0j_{15/2}$ orbital is expected to lie around 1.2~MeV, but yields would be smaller still.  The absolute cross sections have uncertainties of around 30\%. Only relative cross sections were used in the analysis, which are known to better than 5\%.

\begin{table}
\caption{\label{tab1} Excitation energies, tentative $\ell$ and $J^{\pi}$ assignments, and normalized spectroscopic factors $S$ for states assigned to $^{207}$Hg. The normalization is such that the 3$s_{1/2}$ strength is equal to unity.}
\newcommand\T{\rule{0pt}{3ex}}
\newcommand \B{\rule[-1.2ex]{0pt}{0pt}}
\begin{ruledtabular}
\begin{tabular}{ccccc}
 $E$ (keV)\B & $\ell$ & $J^{\pi}$ & $n\ell s$ & $S$  \\
\hline
0\T & 4 & 9/2$^+$ & $1g_{9/2}$ & $0.82(13)$ \\
\T  1195(20) & (2) & (5/2$^+$) & ($2d_{5/2}$)\footnote{The centroid of the $2d_{5/2}$ strength lies at 1500(50)~keV, with $\sum C^2S=1.02(17)$.} & 0.47(9)  \\
\T 1600(45) & (2) & (5/2$^+$) & ($2d_{5/2}$)$^{\rm a}$  & 0.13(2) \\
\T 1810(20)  & (2) & (5/2$^+$) & ($2d_{5/2}$)$^{\rm a}$  & 0.42(7) \\
\T  1960(30) & (0) & (1/2$^+$) & ($3s_{1/2}$) &  $\equiv1.00$  \\
\T  2335(20) & (2) & (3/2$^+$)  & ($2d_{3/2}$) & 1.00(17)   \\
\T\B  2530(20) & (4) & (7/2$^+$) & ($1g_{7/2}$)  &  0.62(12)  \\
\end{tabular}
\end{ruledtabular}
\end{table}

Calculations used to extract spectroscopic factors and predict angular distributions were performed using the distorted-wave Born approximation (DWBA) with the code {\sc Ptolemy}~\cite{Ptolemy78}. The bound-state form factors were taken from Ref.~\cite{Kay13} and optical-model potentials from Refs.~\cite{Koning03} and~\cite{An06} were used. Normalized spectroscopic factors are listed in Table~\ref{tab1}; the uncertainties are dominated by the relative variation due to different choices in the optical-model potentials.

The calculated angular distributions are fitted to the experimental data in Fig.~\ref{fig3}. While at 7.38~MeV/u the distributions are relatively indistinct, tentative $\ell$-value assignments have been made. Each state was fitted with $\ell=0$, 2, and 4 shapes using a $\chi^2$ minimization technique. The $\ell$-value for the resulting best-fit shape was adopted. In the case of the $\ell=0$ state, while having the smallest $\chi^2$, the value was similar to the values for $\ell=2$ and 4, but because of its large cross section, an assignment other than $\ell=0$ would cause serious inconsistencies in the sum rules. The ground state is identified as $\ell=4$ and corresponds to $J^{\pi}=9/2^+$. The states at 1195, 1600, and 1810~keV, are assigned as $\ell=2$, which is similar to the pattern seen in $Z=84$, $^{211}$Po~\cite{Bhatia79}. The core 2$^+$ excitations in $^{206}$Hg and $^{210}$Po are at similar energies of $\sim$1.1~MeV. This can cause fragmentation of the excited states and is likely to be responsible for the three $5/2^+$ fragments and for any other fragmentation of the $2d_{3/2}$ and $1g_{7/2}$ strengths. The details, of course, depend on specific structural considerations. The $3s_{1/2}$ strength is carried by the state at 1960~keV and the $\ell=2$ transfer to the 2335-keV state is presumed to carry the full $2d_{3/2}$ strength, lying just over 0.8~MeV above the $2d_{5/2}$ centroid. The state at 2530~keV is populated via $\ell=4$ transfer, and is assigned $J^{\pi}=7/2^+$, a major fragment of the spin-orbit partner to the ground state.
 
A sum-rule analysis of the spectroscopic factors gives a consistent picture of the assignments. The spectroscopic factors in Table~\ref{tab1} are normalized, arbitrarily, such that the 1960-keV $\ell=0$ transition has $S\equiv1.00$. The summed strength for $1g_{9/2}$, $2d_{5/2}$, $3s_{1/2}$, $2d_{3/2}$, and $1g_{7/2}$ are then 0.82(13), 1.02(14), 1.00, 1.00(17), and 0.62(12), respectively, implying that the bulk of the strength is carried by these states. The observed $1g_{7/2}$ strength is notably lower than others, suggesting fragments of this strength lie at higher excitation energy than was probed here. The ordering of states is consistent with trends seen in $^{209}$Pb and $^{211}$Po, and that expected of single-neutron orbitals outside of the $N=126$ shell closure.

\begin{figure*}
\centering
\includegraphics[scale=0.8]{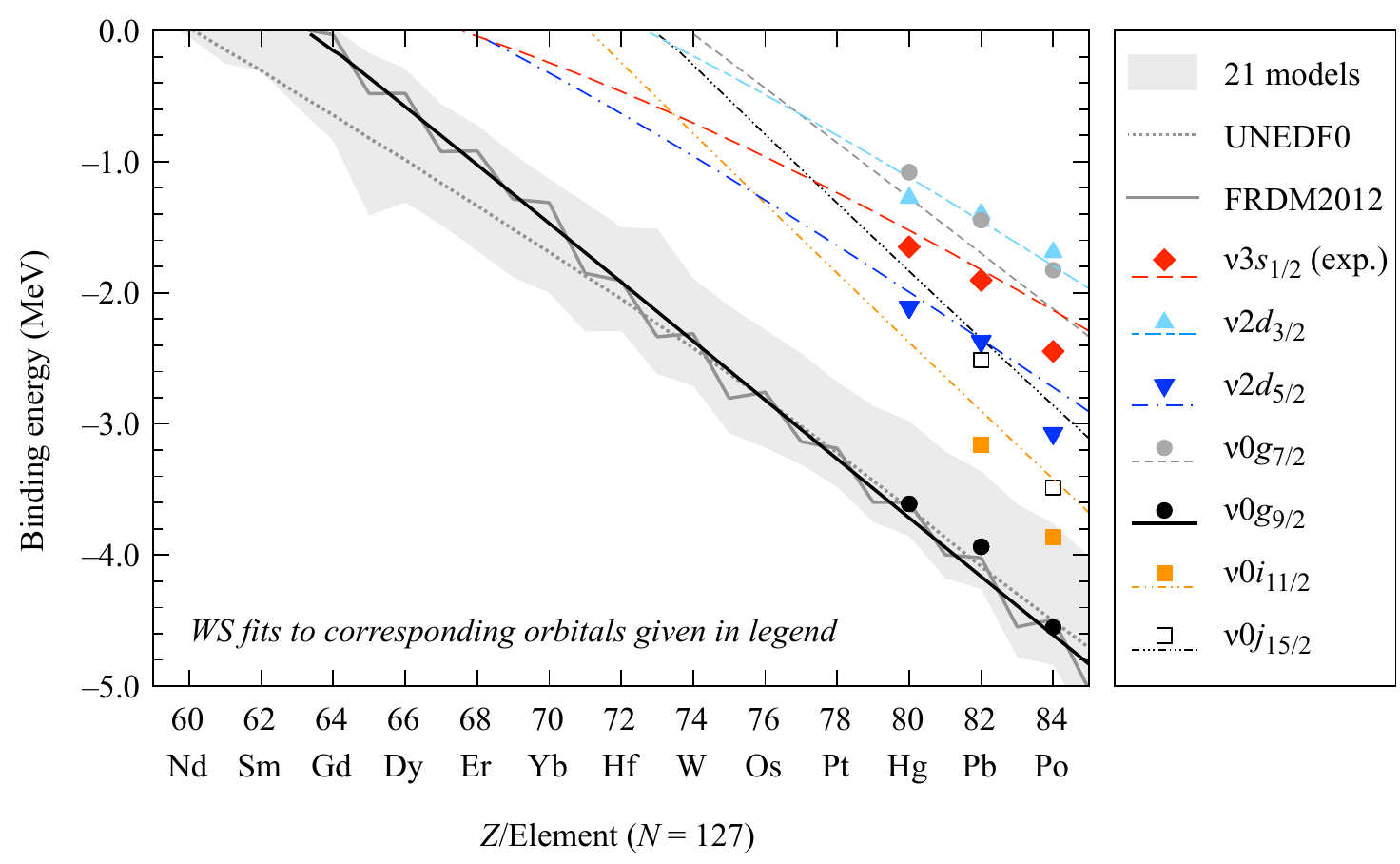}
\caption{\label{fig4} Experimentally determined binding energies of the neutron orbitals at $N=127$ for $^{207}$Hg (this measurement), $^{209}$Pb~\cite{nudat}, and $^{211}$Po~\cite{Bhatia79} compared to Woods-Saxon (WS) calculations of the same orbitals constrained by the Pb and new Hg data. The grey band represents the spread in 21 different models used to calculate $S_n$ for the ground states of  nuclei $N=127$ that are commonly used in $r$-process studies~\cite{Mumpower16a}.  UNEDF0~\cite{unedf0} and FRDM2012~\cite{frdm2012} are highlighted.}
\end{figure*}

Figure~\ref{fig4} shows the experimentally determined binding energies of the $1g_{9/2}$, $2d_{5/2}$, $3s_{1/2}$, $2d_{3/2}$, and $1g_{7/2}$ orbitals for $^{207}$Hg, and those of $^{209}$Pb and $^{211}$Po from ($d$,$p$)-reaction data. For Pb and Po, the $0i_{11/2}$ and $0j_{15/2}$ excitations are also known.
 
The binding energies of the neutron orbitals at $N=127$ have been calculated using a Woods-Saxon potential, with an asymmetry term in the potential depth as defined in Ref.~\cite{Schwierz07}. The potential depth, spin-orbit strength, radius, and diffuseness parameters were determined by fitting them to the experimental data for the five known single-particle centroids in $^{207}$Hg and the seven in $^{209}$Pb. The rms deviation of the fit from the experimental data was $\sim$200~keV and yielded physically sensible parameters.

The binding energies of the ground and excited states, using the parameters derived from the fitting described above, were extrapolated to zero binding along \mbox{$N=127$}. These extrapolations suggest that below Gd ($Z=64$), $N=127$ nuclei are unbound. The uncertainties on this approach are about two units of $Z$. This assumption depends on the robustness of the closed shell at $N=126$. Indeed, above Pb, the $2^+$ energies are essentially constant through to Th ($Z=90$), beyond which there are no data~\cite{nudat}. Heavier Th isotopes away from the closed shell are known to exhibit strong quadrupole deformation, much like the neutron-rich rare-earth nuclei ($Z\lesssim70$), but the shell closure seems to restore the spherical shape. 

For the ground states, results from the Woods-Saxon calculations, constrained by the new experimental data, are compared to results from 21 models used for determining $S_n$ that are commonly used in $r$-process calculations~\cite{Mumpower16a}. These are shown by the grey band in Fig.~\ref{fig4}. Two models close to the extremes with regards to predicting zero binding energy at $N=127$ nuclei are highlighted, UNEDF~\cite{unedf0} and FRDM~\cite{frdm2012}. The latter agrees well with the simple Woods-Saxon extrapolation guided by the experimental data. 

The Woods-Saxon calculations suggest that between Gd ($Z=64$) and Er ($Z=68$), there are no bound {\it excited} states in these $N=127$ nuclei. If there are no bound excited states and only a bound $9/2^+$ ground state it is difficult to see how capture would proceed. The level density in the continuum will be very low and the usual statistical estimates are not applicable. Direct capture can take place if there is an appreciable component of odd-parity phase shifts with $\ell=4\pm1$. But such single-neutron excitations are far from the neutron threshold. Direct capture with higher than an $E1$ multipole is inhibited~\cite{Goriely97}. In odd-$Z$ nuclei, a member of a multiplet coupled to an $h_{11/2}$ proton hole may provide a basis for a neutron capture and subsequent $\beta$ decays. A microscopic description of the bottleneck for the $r$-process will need considerably more data to allow for a more quantitative description. The bottleneck for $N=126$ appears to be more significant than for $N=82$, where for $N>82$ there are odd-parity $p$ orbitals~\cite{Manning19}.

Referring back to the $r$-process path shown in Fig.~\ref{fig1}, the $N=126$ bottleneck starts around $Z=54$, and capture reactions beyond $N=126$, starting in the first few tenths of a second into the $r$-process, occur around $Z=64\pm2$. The Woods-Saxon calculations shown here, constrained to new data at $^{207}$Hg which includes excited states, are broadly in line with this picture. The UNEDF0 and FRDM2012 models used to predict the neutron separation energy also agree with the experimental data at $80<Z<84$, the latter consistent with the Woods-Saxon calculations when extrapolated to low $Z$.

Beyond knowing the ground-state binding energy, knowledge of the excited states in $^{207}$Hg offers a first glimpse of the changes of shell structure of this region, changes that might impact the $r$-process. This work marks the first exploration of the structure of the $N=127$ nucleus $^{207}$Hg, and paves the way for future experimental studies of this region with the new ISS at ISOLDE and at next generation radioactive ion beam facilities about to come online.

This measurement (IS631) was carried out at the ISOLDE facility at CERN. We wish to thank the ISOLDE team for their outstanding commitment to ensuring measurement could occur before the long shutdown at CERN and to delivering the high quality beam. We wish to acknowledge engineering support from Russell Knaack (Argonne National Laboratory) and from the team at Daresbury Laboratory, and target-development support from Andreea Mitu (IFIN-HH). This material is based upon work supported by the U.S.\ Department of Energy, Office of Science, Office of Nuclear Physics, under Contract Number DE-AC02-06CH11357 (ANL), the UK Science and Technology Facilities Council (Grant Numbers ST/P004598/1, ST/N002563/1, ST/M00161X/1 [Liverpool]; ST/P004423/1 [Manchester]; ST/P005314/1 [Surrey]; and ST/P005101/1 [West of Scotland]; and the ISOL-SRS Grant [Daresbury]), the European Union's Horizon 2020 Framework research and innovation program under grant agreement No. 654002 (ENSAR2) and the Marie Skłodowska-Curie grant agreement No. 665779, and from the Research Foundation Flanders (FWO, Belgium) under the Big Science project G0C28.13, and the European Research Council under the European Union’s Seventh Framework Programme (FP7/2007-2013) / ERC grant agreement No. 617156. MM was supported by the US Department of Energy through the Los Alamos National Laboratory. Los Alamos National Laboratory is operated by Triad National Security, LLC, for the National Nuclear Security Administration of U.S.\ Department of Energy (Contract Number 89233218CNA000001). MM was also supported by the Laboratory Directed Research and Development program of Los Alamos National Laboratory under project number 20190021DR.



\begin{references}
\bibitem{fits} Fits were done using Gaussian peaks of fixed full width at half maximum on a linear background.
\end{references}
\end{document}